\documentclass[preprint,12pt]{elsarticle}

\usepackage{amssymb}
\usepackage{latexsym,bm}
\usepackage{amsmath}
\usepackage{caption2}

\begin{document}

\begin{frontmatter}



\title{Pre-neutron-emission mass distributions for reaction $^{232}$Th(n, f) up to 60 MeV}


\author[label1]{Xiaojun Sun\corref{cor1}}
\cortext[cor1]{Corresponding author.}
\ead{sxj0212@gxnu.edu.cn}

\author[label1]{Chenghua Pan}
\author[label2]{Chenggang Yu}
\author[label1]{Yongxu Yang}

\author[label1]{Ning Wang\corref{cor2}}

\address[label1]{Department of physics, Guangxi Normal University, Guilin 541004, People's Republic of China}
\address[label2]{Shanghai Institute of Applied Physics, Chinese Academy of Sciences, Shanghai 201800, People's Republic of China}

%

\begin{abstract}
The pre-neutron-emission mass distributions for reaction $^{232}$Th(n, f) up to 60 MeV are systematically studied with an empirical fission potential model. The energy dependences of the peaks and valleys of the pre-neutron-emission mass distributions are described by the exponential expressions based on the newly measured data. The energy dependence of evaporation neutrons before scission, which plays a crucial role for the reasonable description of the mass distribution, is also considered. Both the double-humped and triple-humped shape of the measured pre-neutron-emission mass distributions for reaction $^{232}$Th(n, f) are reasonably well reproduced at incident energies up to 60 MeV. The mass distributions at unmeasured energies and the critical energies at which the humped pre-neutron-emission mass distributions are transformed into each other are also predicted.
\end{abstract}

\begin{keyword}

$^{232}$Th(n, f); Pre-neutron-emission mass distribution; evaporation neutron; fission potential

\PACS 24.75.+i; 25.85.Ec
\end{keyword}

\end{frontmatter}


\section{Introduction}
\label{sect1}

The pre-neutron-emission mass distribution is one of the most important characteristics of the nuclear fission process. In nuclear energy applications, the composition of fission products must be known because they are accumulated during the operation of a nuclear reactor and influence physical and chemical properties of nuclear fuel. It is necessary to accurately describe and predict fission yields at different energies for a successful nuclear fission model. Due to the complex of fission process, it is still difficult to microscopically describe the mass splitting for neutron-induced fission at low and intermediate energies \cite{X. Sun2012,sun2013,Ryzhov2011}, though it is believed that the formation of the fission fragment-mass distribution is closely connected with the potential energy surface in deformation space \cite{J. Randrup2011prl,J. Randrup2011prc}.
Nowadays, there is an increasing interest in studying neutron-induced fission of actinides at intermediate energies. It is motivated by nuclear data needs for new applications such as
accelerator-driven system, thorium-based fuel cycle, and the next generation of exotic beam facilities. However, the data on fission fragment mass yields are scarce at neutron energies above 10 MeV, especially for reaction $^{232}$Th(n, f). Recently, the reaction $^{232}$Th(n, f) at intermediate energies was measured by V.D. Simutkin group \cite{Ryzhov2011,V. Simutkin2011,V. Simutkin2013}. Theoretical calculations for the pre-neutron-emission mass distributions is of great importance for understanding the fission process and for describing the measured yields of the fission products.

Compared with low-energy fission, neutron-induced fission at intermediate energies is more complicated due to the pre-equilibrium particle emission and neutron evaporation. There are several codes, such as MCFX \cite{Grudzevich2007}, TALYS \cite{Koning2008}, UKFY4.1 \cite{Mill2008}, GEF2012/2.4 \cite{Schmidt2012} and PYF \cite{Gorodisskiy2008}, which can calculate the pre-neutron-emission mass distributions for reaction $^{232}$Th(n, f) at low or intermediate energies. Generally, the agreement between the experimental data and the mode calculations mentioned above is good for $^{238}$U but worse for $^{232}$Th at intermediate energies \cite{Ryzhov2011,V. Simutkin2011,V. Simutkin2013}.
It can be obviously seen from the experimental data \cite{Ryzhov2011,V. Simutkin2011,V. Simutkin2013,L. E. Glendenin1980,H. Naika2013} that the pre-neutron-emission mass distributions for reaction $^{232}$Th(n, f) gradually change from double-humped to triple-humped shape with increasing the incident energies. In this work, we attempt to describe quantitatively both the double-humped and triple-humped pre-neutron-emission mass distributions for reaction $^{232}$Th(n, f) up to 60 MeV with an empirical fission potential.

The empirical fission potential and its parameters are introduced in Section \ref{sect2}. The calculated results for the pre-neutron mass distributions from the model are shown in Section \ref{sect3}. In section \ref{sect4}, we give a brief summary.

\section{Fission potential and its parameters}
\label{sect2}

\begin{figure}
\centering
\includegraphics[width=8cm,angle=0]{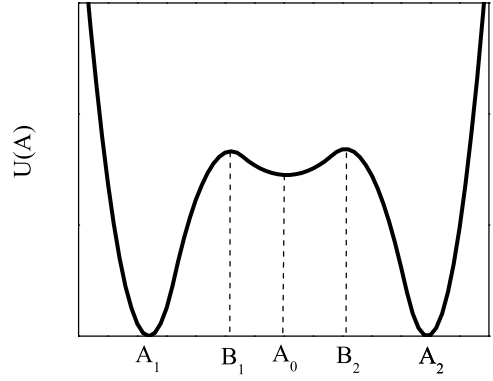}
\caption{Schematic representation of the fission potential. }\label{potential}
\end{figure}

Assuming that a compound nucleus separates into a pair of daughter nuclei in the fission process, so the pre-neutron-emission mass distributions of intermediate energies neutron-induced $^{232}$Th fission can be approximately described by a simplified fission potential $ U(A)$,
\begin{equation}\label{eq1}
 P{(A)}=C \exp[-U(A)].
\end{equation}
Where $ C $ is the normalization constant, and the variable $ A $ denotes the mass number of the primary fragment. Considering the triple-humped mass distributions of intermediate energies neutron-induced $^{232}$Th fission, we describe the phenomenological fission potential $U(A)$ by using five harmonic-oscillator functions (see Fig. \ref{potential}), i.e.,
\begin{equation}\label{eq2}
U(A)=
\begin{cases}
U_1(A-A_1)^2                 & A\leq a, \\
-V_1(A-B_1)^2+R_1            &a\leq A \leq b, \\
U_2(A-A_0)^2+R_0            &b\leq A \leq c, \\
-V_2(A-B_2)^2+R_2            &c\leq A \leq d, \\
U_3(A-A_2)^2                 & A\geq d.
\end{cases}
\end{equation}
Where $ A_1$ and $ A_2 $ are the positions for the peaks of the light and heavy fragments of the pre-neutron-emission mass distributions, and $A_0$ is the position of the symmetrical fission. $B_1$ and $B_2$ are the positions for the valleys of the light and heavy fragments of the pre-neutron-emission mass distributions, respectively.
Considering that the fission potential is a smooth function, the coefficients in Eq. (\ref{eq2}) can be derived as
\begin{equation}\label{eq3}
\begin{split}
 U_1&=\frac{R_1}{(A_1-B_1)(A_1-a)}, \\
 V_1&=\frac{R_1}{(A_1-B_1)(a-B_1)}, \\
 U_2&=-\frac{V_1(b-B_1)}{b-A_0}, \\
 V_2&=-\frac{U_2(c-A_0)}{c-B_2}, \\
 U_3&=-\frac{V_2(d-B_2)}{d-A_2}, \\
 b&=B_1+\frac{R_1-R_0}{V_1(A_0-B_1)}, \\
 c&=A_0+\frac{R_0-R_2}{U_2(A_0-B_2)}, \\
 d&=B_2+\frac{R_2}{V_2(A_2-B_2)}. \\
 \end{split}
\end{equation}
The potential parameters $a, R_1, R_2$ and $R_0$ will be discussed later.

The total mass distributions of the binary fission fragments should be normalized to 200\%. The normalization constant $ C $ can therefore be analytically expressed as
\begin{equation}\label{eq4}
\begin{split}
 C&=\frac{200\%}{\int_0^\infty U(A)dA} \\
 &=\frac{200\%}{I_0+I_1+I_2+I_3+I_4},
 \end{split}
\end{equation}
with
\begin{equation}\label{eq5}
\begin{split}
 I_0&=\frac{\sqrt{\pi}}{2\sqrt{U_1}}\{\textmd{erf}[(a-A_1)\sqrt{U_1}]+\textmd{erf}[A_1\sqrt{U_1}]\}, \\
 I_1&=\frac{\sqrt{\pi}e^{-R_1}}{2\sqrt{V_1}}\{-\textmd{erfi}[(a-B_1)\sqrt{V_1}]+\textmd{erfi}[(b-B_1)\sqrt{V_1}]\}, \\
 I_2&=\frac{\sqrt{\pi}e^{-R_0}}{2\sqrt{U_2}}\{\textmd{erf}[(A_0-b)\sqrt{U_2}]-\textmd{erf}[(A_0-c)\sqrt{U_2}]\}, \\
 I_3&=\frac{\sqrt{\pi}e^{-R_2}}{2\sqrt{V_2}}\{\textmd{erfi}[(B_2-c)\sqrt{V_2}]-\textmd{erfi}[(B_2-d)\sqrt{V_2}]\}, \\
 I_4&=\frac{\sqrt{\pi}}{2\sqrt{U_3}}\{1+\textmd{erf}[(A_2-d)\sqrt{U_3}]\}, \\
 \end{split}
\end{equation}
where erf(x) and erfi(x) denote the error function and imaginary error function, respectively. We also assume that $P(A_1) = P(A_2)$ and $P(B_1) = P(B_2)$, i.e., for the pre-neutron-emission mass distributions, the heights of the peaks and valleys of the light fragments equal those of the heavy fragments. So the parameter $a$ can be uniquely determined by the experimental data $P(A_0), P(A_1), P(B_1)$ and the normalization constant $C$.

In addition, the potential parameters $R_1, R_2$ and $R_0$ can be derived from the height of peaks and valleys, i.e.,
\begin{equation}\label{eq6}
\begin{split}
R_1=R_2=\ln\frac{P{(A_1)}}{P{(B_1)}},\\
R_0=\ln\frac{P{(A_1)}}{P{(A_0)}}.
\end{split}
\end{equation}

A particular attention should be payed to that these position parameters are closely relative to the evaporation neutrons before scission at different incident energies.
For reaction $^{232}$Th(n, f) at low incident energies ($E_n < 6.0$ MeV), the pre-neutron-emission mass distributions show a double-damped shape, and the fission potential Eq. (\ref{eq1}) will be simplified as the three harmonic-oscillator functions \cite{X. Sun2012}.
The positions $A_1$ and $A_2$ are obtained from the nucleus-nucleus
driving potential of the fissile nucleus $^{233}$Th \cite{X. Sun2012,Liu2006,Wang2009}.
With increasing the incident neutron energy, the excitation energy of the compound nucleus will become higher, and a few neutrons will be
evaporated before scission. The number of the evaporation neutron $\tilde{n}(E_n)$ is empirically expressed as

\begin{equation}\label{eq7}
\begin{cases}
\displaystyle \tilde{n}(E_n)=0, &E_{th} \leq E_n \leq 6.0~\textmd{MeV}, \\
\displaystyle \tilde{n}(E_n)=1, &6.0 < E_n \leq 14.0~\textmd{MeV}, \\
\displaystyle \tilde{n}(E_n)=2, &14.0 < E_n \leq 21.0~\textmd{MeV}, \\
\displaystyle \tilde{n}(E_n)=3, &21.0 < E_n \leq 31.0~\textmd{MeV}, \\
\displaystyle \tilde{n}(E_n)=4, &31.0 < E_n \leq 52.0~\textmd{MeV}, \\
\displaystyle \tilde{n}(E_n)=5, &52.0 < E_n \leq 60.0~\textmd{MeV}. \\
\end{cases}
\end{equation}
Where, $E_{th}$ is the threshold energy for $^{232}$Th(n, f) reaction. It is assumed that a compound nucleus (mass number is $A_{CN}$) after evaporating neutrons separates into a pair of daughter fragments in the fission process, and the mass number of the fissile nucleus is $A_{FN}$ at different incident energy regions. For neutron induced $^{232}$Th fission, the fissile nuclei are $^{233}$Th, $^{232}$Th,  $^{231}$Th, $^{230}$Th, $^{229}$Th and $^{228}$Th corresponding the reaction channels (n, f), (n, nf), (n, 2nf), (n, 3nf), (n, 4nf) and (n, 5nf), respectively. The position $A_2$ of the heavy fragment peaks, as well as $A_1$ of the light fragment peaks and $A_0 $ of the symmetrical fission, should be modified as
\begin{equation}\label{eq8}
\left\{\begin{array}{llllll}
        \displaystyle A_{FN}&=A_{CN}-\tilde{n}(E_n),    \\
        \displaystyle A_1&=A_{CN}-A_2^{g.s.},     \\
        \displaystyle A_2&=A_2^{g.s.}-\tilde{n}(E_n),      \\
        \displaystyle A_0&=A_{FN}/2.     \\
       \end{array}\right.
\end{equation}
Where $A_2^{g.s.}\simeq 140$ denotes the lowest position of the driving potential derived from the ground state of the fissile nucleus \cite{X. Sun2012}.

\begin{figure}
\centering
\includegraphics[width=8cm,angle=0]{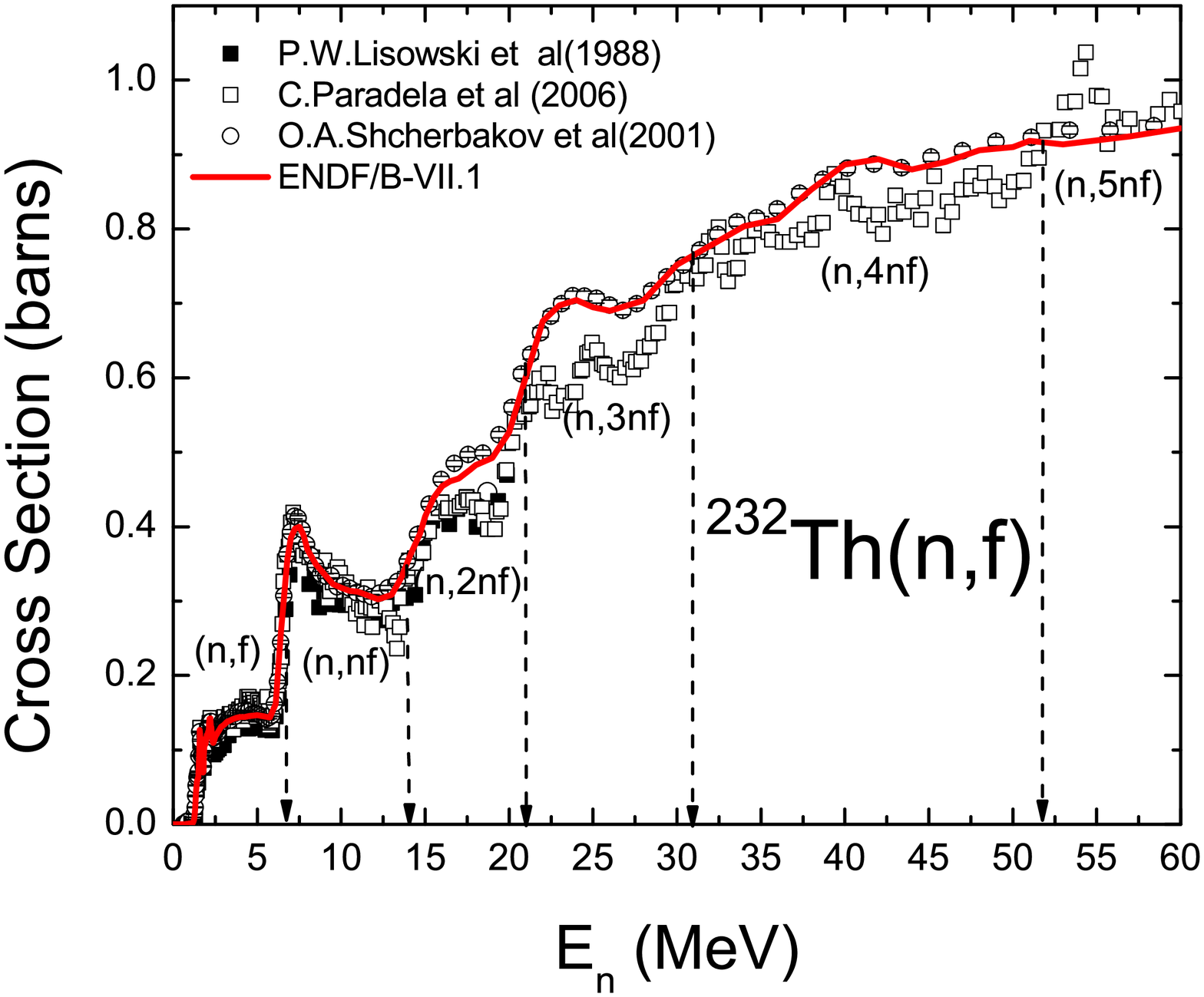}
\caption{ (Color online) Fission cross section of reaction $^{232}$Th(n, f) for incident neutron energies from threshold energy to 60 MeV. The experimental data are obtained from Refs. \cite{P.W.1988} (full squares), \cite{Paradela2006} (empty squares) and \cite{Shcherbakov2001}(empty circles), respectively. The solid curve denotes the evaluated results of ENDF/B-VII.1, and the dash lines denote the incident energy intervals corresponding to the different multi-chance fission channels as labeled (n, f), (n, nf), (n, 2nf), (n,3nf), (n, 4nf) and (n, 5nf), respectively. }\label{CS}
\centering
\includegraphics[width=8.5cm,angle=0]{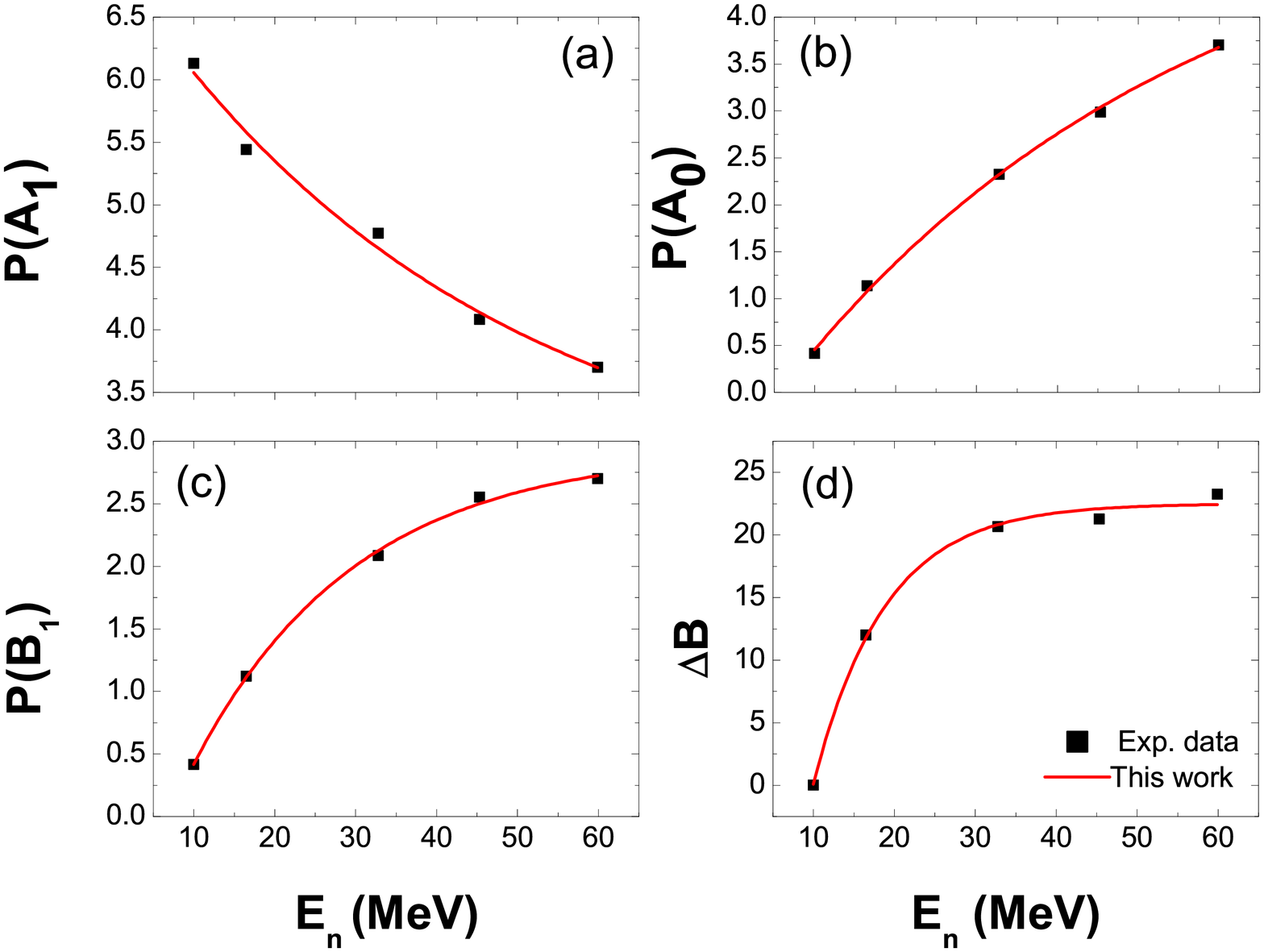}
\caption{ (Color online) Heights of pre-neutron-emission mass distributions for (a) light fragment peak $P(A_1)$, (b) symmetric fission point $P(A_0)$, (c) light fragment valley $P(B_1)$ and (d) the width between valleys $\triangle B$ in fission reaction $^{232}$Th(n, f) as a function of incident neutron energy $E_n$. The experimental data are derived from Refs. \cite{Ryzhov2011,V. Simutkin2011,V. Simutkin2013}, and the curves denote the results of Eq. (\ref{eq9}) and Eq. (\ref{eq11}). }\label{PA}
\end{figure}

Additionally, the different incident energy intervals of Eq. (\ref{eq7}) corresponding the evaporation neutron numbers are consistent with the fission cross sections for reaction n+$^{232}$Th as shown in Fig. \ref{CS}. The scattering dots denote the experimental data derived from Refs. \cite{P.W.1988,Paradela2006,Shcherbakov2001}, and the solid curve denotes the evaluated results of ENDF/B-VII.1. The dash lines denote the incident energy regions corresponding to the different multi-chance fission channels as labeled  (n, f), (n, nf), (n, 2nf), (n, 3nf), (n, 4nf) and (n, 5nf), respectively.

For the triple-humped pre-neutron-emission mass distributions, the positions of the valleys $B_1$ and $B_2$ shown in Eq. (\ref{eq2}) play an important role.
Based on the quasi-monoenergetic experimental data \cite{Ryzhov2011,V. Simutkin2011,V. Simutkin2013}, one can see that the width between the valleys $\Delta B=B_2-B_1$ increases with the incident energies, and its energy dependence can be reasonably well described with a parameterized formula
\begin{equation}\label{eq9}
\Delta B=22.509-69.774\exp(-0.114E_n).
\end{equation}
So the positions of the valleys $B_1$ and $B_2$ should be modified as
\begin{equation}\label{eq10}
\begin{split}
 B_1&=A_0-\Delta B/2, \\
 B_2&=A_0+\Delta B/2.
 \end{split}
\end{equation}

Based on the experimental data \cite{Ryzhov2011,V. Simutkin2011,V. Simutkin2013} of the reaction $^{232}$Th(n, f) up to 60 MeV, the energy dependence of heights of the peaks $P(A_1)$, $P(A_0)$ and the valley $P(B_1)$ is also exponentially expressed as
\begin{equation}\label{eq11}
\begin{split}
P(A_1)&=2.570+4.371\exp(-0.023E_n),  \\
P(A_0)&=5.543-6.220\exp(-0.020E_n),  \\
P(B_1)&=2.934-4.139\exp(-0.050E_n).
\end{split}
\end{equation}

One can see that the values of $P(A_1)$ , $P(A_0)$ and $P(B_1)$ exponentially change with the incident energies in general, which could provide some useful information at unmeasured energies.

\section{Results and discussion}
\label{sect3}

\begin{figure}
\centering
\includegraphics[width=8cm,angle=0]{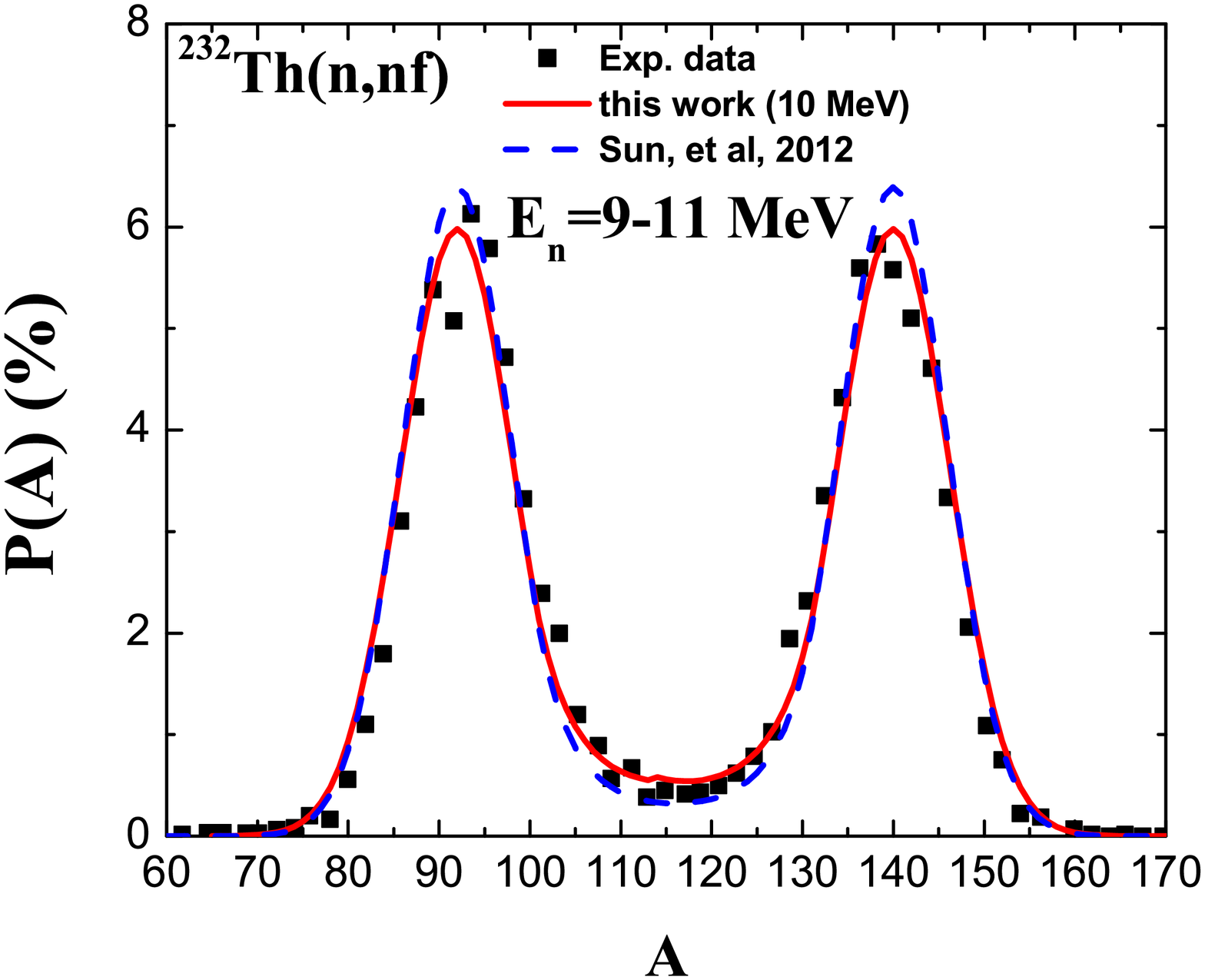}
\caption{ (Color online) Pre-neutron-emission mass distributions for reaction $^{232}$Th(n, nf). The scattered symbols denote the experimental data in incident energy interval 9-11 MeV, which are taken from Ref. \cite{V. Simutkin2011,V. Simutkin2013} (squares). The red solid curve denotes the calculated results at $E_n$=10 MeV, and the blue dash curve does the results using the parameters of Ref. \cite{X. Sun2012} at the same incident energy. }\label{PA11}
\centering
\includegraphics[width=8cm,angle=0]{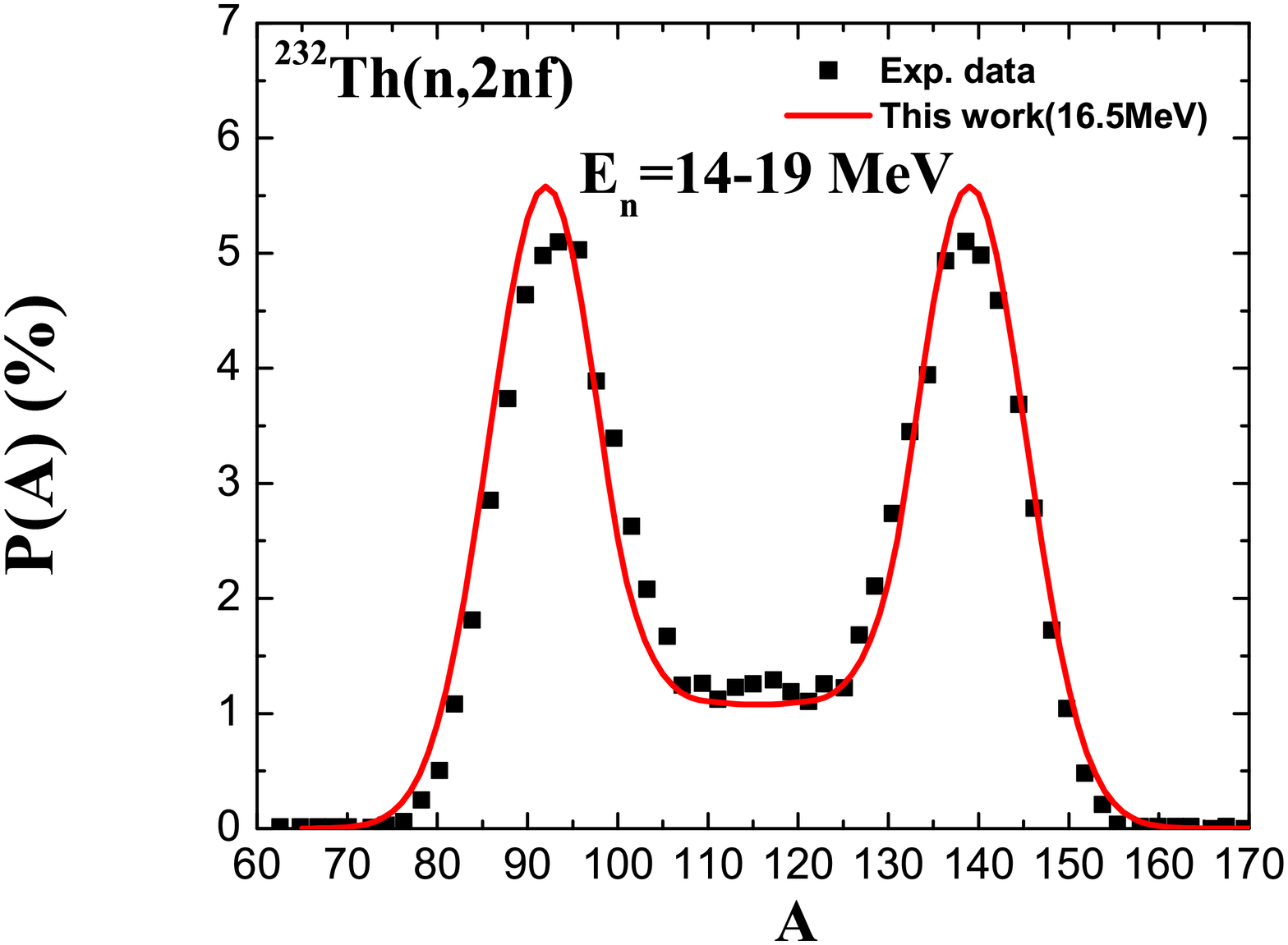}
\caption{(Color online) Pre-neutron-emission mass distributions for reaction $^{232}$Th(n, 2nf). The scattered symbols denote the experimental data in incident energy interval 14-19 MeV \cite{V. Simutkin2011,V. Simutkin2013}. The red solid curve denotes the calculated results at $E_n$=16.5 MeV. }\label{PA16.5}
\end{figure}

\begin{figure}
\centering
\includegraphics[width=8cm,angle=0]{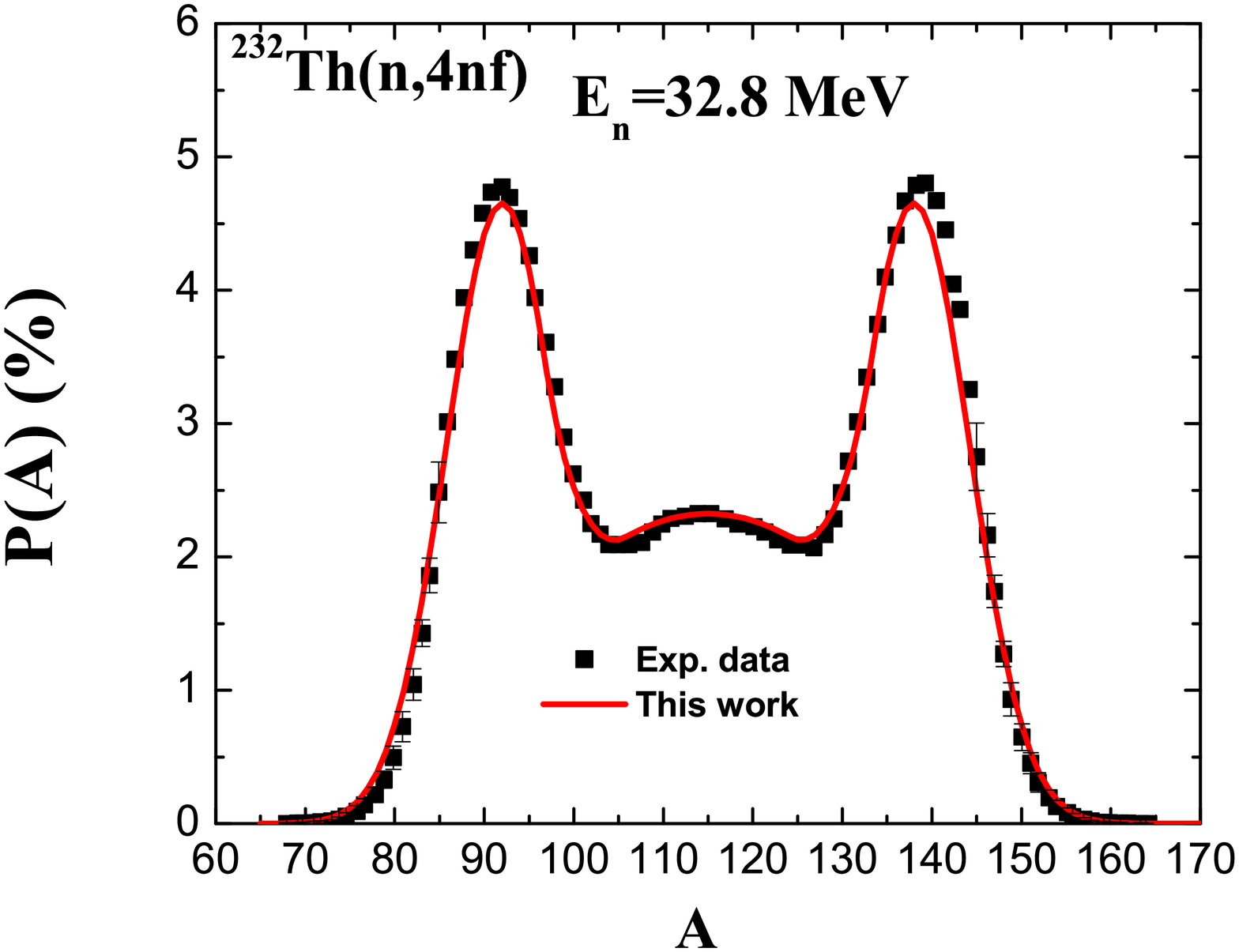}
\caption{(Color online) Pre-neutron-emission mass distributions for reaction $^{232}$Th(n, 4nf). The scattered symbols denote the experimental data \cite{Ryzhov2011,V. Simutkin2011}. The red solid curve denotes the calculated results at $E_n$=32.8 MeV. }\label{PA32.8}
%
\centering
\includegraphics[width=8cm,angle=0]{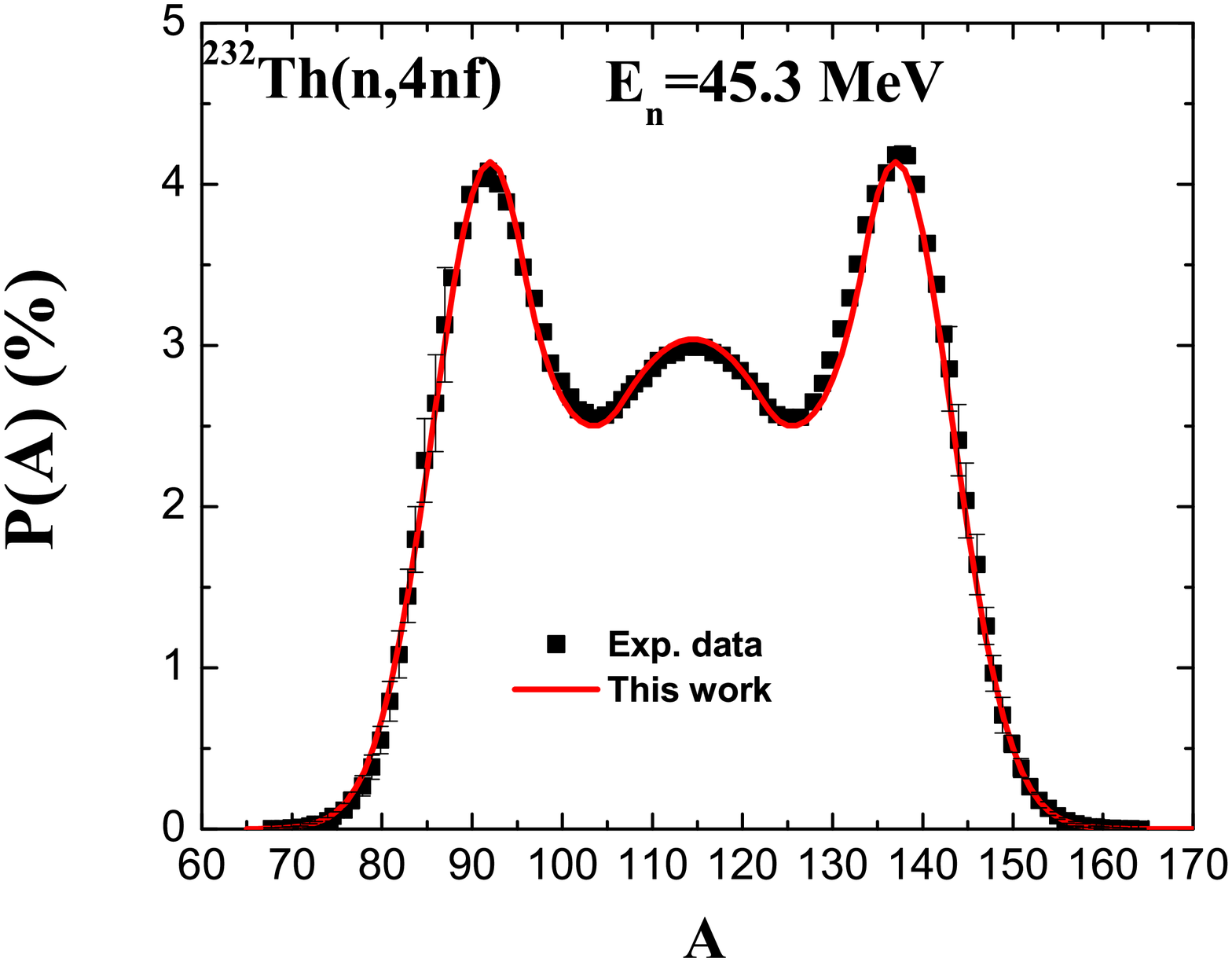}
\caption{ (Color online) The same as Fig. \ref{PA32.8}, but for reaction $^{232}$Th(n, 4nf) at $E_n$=45.3 MeV. }\label{PA45.3}
\end{figure}

\begin{figure}
\centering
\includegraphics[width=8cm,angle=0]{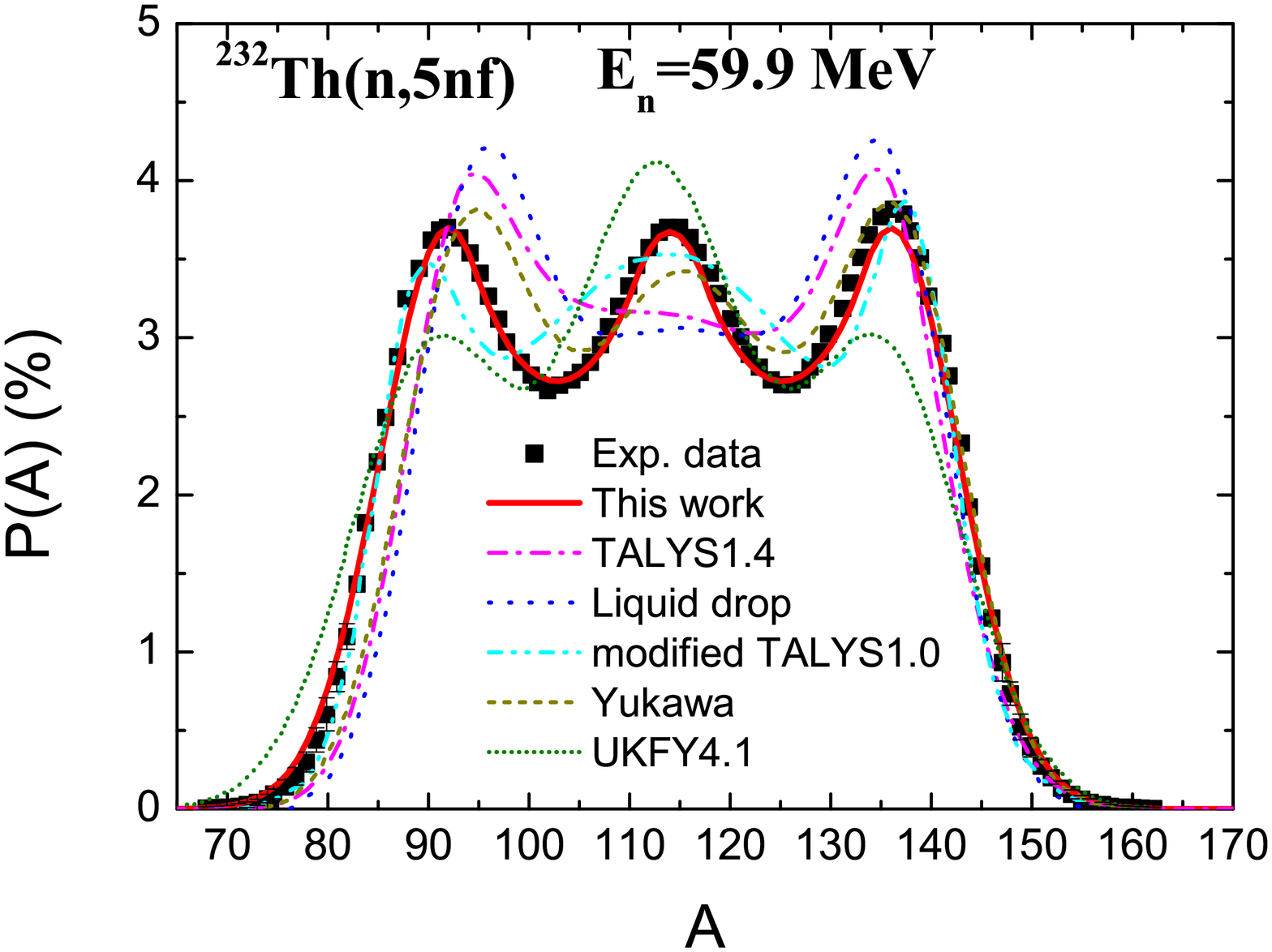}
\caption{ (Color online) Pre-neutron-emission mass distributions for reaction $^{232}$Th(n, 5nf). The scattered symbols denote the experimental data \cite{Ryzhov2011,V. Simutkin2011}. The red solid curve denotes the calculated results of this work at $E_n$=59.9 MeV. The calculation results from different methods are also presented for comparison: TALYS1.4 code \cite{Koning2008,Koning2001} (magenta dash dot curve), the liquid drop model \cite{Seeger1976} (blue dot curve), modified TALYS1.0 code \cite{V. Simutkin2011} (cyan dash dot dot curve), Yukawa \cite{Dobrowolski2007} (dark yellow short dash curve) and UKFY4.1 library \cite{Mill2008} (olive short dot curve).}\label{PA59.9}
%
\centering
\includegraphics[width=11cm,angle=0]{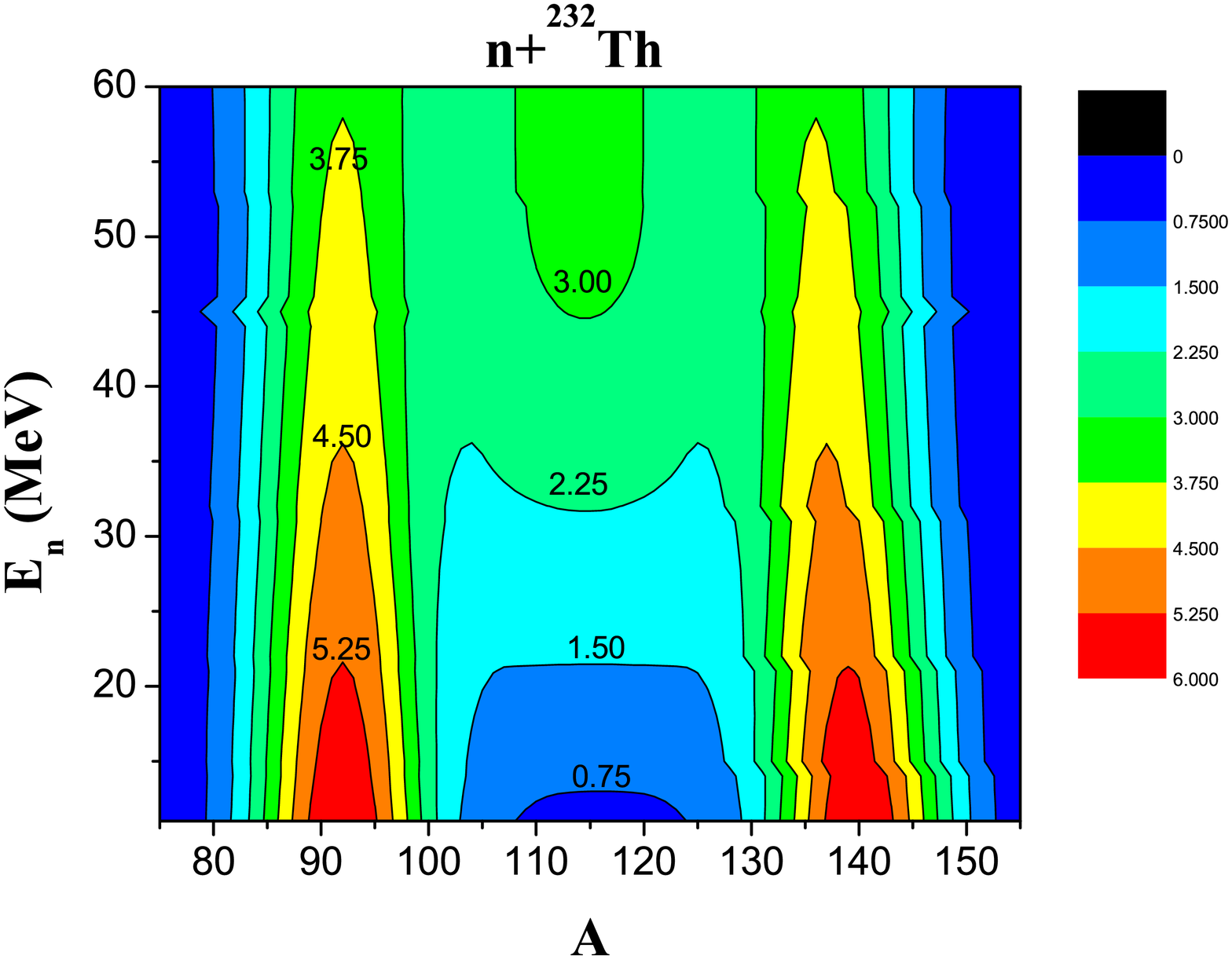}
\caption{(Color online) Predicted pre-neutron-emission mass distributions {(}\%{)} at
incident energies from 10 MeV to 60 MeV for reaction $^{232}$Th(n, f).}\label{PA60}
\end{figure}

In this work, the evaporation neutrons at different incident energy intervals are empirically derived, and are additionally consistent with the fission cross sections in multi-chance fission channels as shown in Fig. \ref{CS}. In terms of the driving potential, the position of the light fragment peak $A_1$ of the pre-neutron-emission mass distributions for reaction $^{232}$Th(n, f) is fixed. The energy dependence of the positions of the heavy fragment peaks $A_2$ and the symmetric fission point $A_0$ are uniquely determined after considering the evaporation neutron numbers, as expressed Eq. (\ref{eq8}). Based on the experimental data, the energy dependence of the positions $B_1$ and $B_2$ of the fragment valleys, as well as the energy dependence of the heights of the peaks $P(A_1)$, $P(A_0)$ and the valley $P(B_1)$, are also reasonably described by Eqs. (\ref{eq9})-(\ref{eq11}). Therefore, the parameters of the fission potential are uniquely determined by using Eqs. (\ref{eq2})-(\ref{eq6}). The calculated pre-neutron-emission mass distributions for reaction $^{232}$Th(n, f) are shown in Figs. \ref{PA11}-\ref{PA60}.

A particular attention should be payed to that there is a theoretically critical energy point while the height of the light fragment valley $P(B_1)$ almost equals the value of the symmetric fission point $P(A_0)$ with decreasing the incident energies. So the phenomenological fission potential Eq. (\ref{eq2}) consisting of five harmonic-oscillator functions would be transformed into the expression of three harmonic-oscillator functions as shown in Ref. \cite{X. Sun2012}. Using the parameters mentioned above, the pre-neutron-emission mass distributions for reaction $^{232}$Th(n, f) as a function of the fragment mass number $A$ at incident neutron energy $E_n$=10 MeV are shown in Fig. \ref{PA11} (red solid curve). In this figure, the scattered symbols denote the experimental data in incident energy interval 9-11 MeV, which are taken from Ref. \cite{V. Simutkin2011,V. Simutkin2013}. For comparison, we also give the calculated results using the potential parameters of Ref. \cite{X. Sun2012} at $E_n$=10 MeV (blue dash curve). From Fig. \ref{PA11}, one can see that the results of the mass distributions are slightly improved with the new parameters comparing with those in Ref. \cite{X. Sun2012}. In Ref. \cite{X. Sun2012}, the energy dependence of peaks and valleys is simply assumed as a linear function of energy for fission at low incident energies ($E_n < 6$) MeV. In addition, the calculated pre-neutron-emission mass distributions at different incident energies and the experimental data are also shown in Figs. \ref{PA16.5} - \ref{PA59.9}. Furthermore, the pre-neutron-emission mass distributions at different incident energies from 10 MeV to 60 MeV for reaction $^{232}$Th(n, f) are also predicted (see Fig. \ref{PA60}). This figure clearly shows that the incident energy interval from double-dumped to triple-dumped shape locals 20 MeV - 30 MeV, and the symmetric fission component increases with incident neutron energy.

In Fig. \ref{PA59.9}, we also compare the experimental data at $E_n$=59.9 MeV with the model calculations from different methods. The red solid, magenta dash dot, blue dot, cyan dash dot dot, dark yellow short dash, olive short dot curves denote the results of this work, TALYS1.4 code \cite{Koning2008,Koning2001}, the liquid drop model \cite{Seeger1976}, modified TALYS1.0 code \cite{V. Simutkin2011}, Yukawa \cite{Dobrowolski2007} and UKFY4.1 library \cite{Mill2008}, respectively. The detailed descriptions of these model calculations are given in Ref. \cite{Ryzhov2011}. Compared with these model calculations, one can see that the results of this work reproduce the experimental data better.

In addition, from Fig. \ref{PA11}- Fig. \ref{PA60}, one can see the double-dumped and triple-dumped shapes of the pre-neutron-emission mass distributions. The triple-dumped pre-neutron-emission mass distributions gradually change to the double-dumped shape with decreasing the incident energies. For reaction $^{232}$Th(n, f), this critical incident energy is $\sim 23$ MeV through comparing the heights of the valley $P(B_0)$ and the symmetric fission point $P(A_0)$ as expressed Eq. (\ref{eq11}). In addition, we find that at energy $E_n \geq 110$ MeV the pre-neutron-emission mass distributions for reaction $^{232}$Th(n, f) change to a single-dumped shape, i.e., the symmetric fission component dominates in the fission process.

\section{Summary}
\label{sect4}
In this work, an empirical fission potential model is proposed to quantitatively describing the pre-neutron-emission mass distributions for reaction $^{232}$Th(n, f) up to 60 MeV. The energy dependences of evaporation neutrons before scission, peaks and valleys of the pre-neutron-emission mass distributions are described by the exponential expressions based on the newly measured data. Both the double-humped and triple-humped shape of the measured pre-neutron-emission mass distributions are reasonably well reproduced from low to intermediate incident energies. The pre-neutron-emission mass distributions at unmeasured energies are also predicted with this approach up to 60 MeV. We compare the calculated results using the previous method and the parameters \cite{X. Sun2012}, and the results of this work are slightly improved at low incident energies. We also compared the experimental data at intermediate energy with the model calculations from different methods, and one can see that the results of this work reproduce the experimental data better. Based on this empirical fission potential model and parameters, we also predict that the critical incident energies are $\sim 23$ MeV from double-dumped to triple-dumped shape, and $\geq 110$ MeV from triple-dumped to single-humped shape of the pre-neutron-emission mass distributions for reaction $^{232}$Th(n, f). This
investigation is helpful for further describing the yields of
the fission products. In addition, a more microscopic description of the potential parameters should be
further investigated. The study of these aspects is underway.
\\

\textbf{Acknowledgements}

We thank Dr. Li Ou, Dr. Min Liu and Dr. Yun Guo for some valuable suggestions. This work was
supported by Guangxi University Science and Technology Research Projects (Grant No. 2013ZD007), GuangXi Natural Science Foundation (Grant No. 2012GXNSFAA053008), National Natural Science Foundation of China (Grants No. 11265004) and the Th-based Molten Salt Reactor Power System of the Strategic Pioneer Science and Technology Projects from the Chinese Academy of Sciences.






\begin{thebibliography}{100}

\bibitem{X. Sun2012}X.J. Sun, C.G. Yu and N. Wang, Phys. Rev. C 85 (2012) 014613.

\bibitem{sun2013}X.J. Sun, C.G. Yu, N. Wang, Y.X. Yang, Pre-neutron-emission mass distributions for reaction $^{238}$U(n, f) up to 60 MeV, arXiv preprint arXiv:1310.2999, (2013).

\bibitem{Ryzhov2011}I.V. Ryzhov, S.G. Yavshits, G.A. Tutin, N.V. Kovalev, A.V. Saulski, N.A. Kudryashev, M.S. Onegin, L.A. Vaishnene, Yu.A. Gavrikov, O.T. Grudzevich, J.P. Meulders, R. Prieels, Phys. Rev. C 83 (2011) 054603.

\bibitem{J. Randrup2011prl}J. Randrup, P. M$\ddot{o}$ller, Phys. Rev. L 106 (2011) 132503.

\bibitem{J. Randrup2011prc}J. Randrup, P. M$\ddot{o}$oller, A.J. Sierk, Fission-fragment mass distributions from strongly damped shape evolution. Phys. Rev. C 84 (2011) 034613.

\bibitem{V. Simutkin2011} V.Simutkin, Fragment mass distributions in neutron-induced fission of $^{232}$Th and $^{238}$U from 10 to 60 MeV. Ph. D. thesis, Acta Universitatis Upsaliensis, 2011.

\bibitem{V. Simutkin2013}V.D. Simutkin, S. Pomp, J. Blomgren, M. $\ddot{O}$sterlund, R. Bevilacqua, I.V. Ryzhov, G.A. Tutin, S.G. Yavshits, L.A. Vaishnene, M.S. Onegin, J.P. Meulders, R. Prieels, Experimental Neutron-Induced Fission Fragment Mass Yields of $^{232}$Th and $^{238}$U at Energies from 10 to 33 MeV. arXiv preprint arXiv:1304.2316, (2013).

\bibitem{Grudzevich2007}O. Grudzevich, S. Yavshits, in Proceedings of the International Conference on Nuclear Data for Science and Technology, Nice, 2007, Vol. 2, p. 1213.

\bibitem{Koning2008}A. Koning, S. Hilaire, M. Dujvestijn, Proceedings of the International Conference on Nuclear Data for Science and Technology, Nice, 211 (2008).

\bibitem{Mill2008}R. Mills, JEF/DOC-1232 and UKNSF, 2008, P227.

\bibitem{Schmidt2012}K.-H. Schmidt, B. Jurado JEFF/DOC report, 1423(2012).

\bibitem{Gorodisskiy2008}D.M. Gorodisskiy, K.V. Kovalchuk, S.I. Mulgin, A.Ya. Rusanov, S.V. Zhdanov, Ann. Nucl. Energy, 35 (2008) 238.

\bibitem{L. E. Glendenin1980}L.E. Glendenin, J.E. Gindler, I. Ahmad, D.J. Henderson, J.W. Meadows. Phys. Rev. C 22 (1980) 152.

\bibitem{H. Naika2013}H. Naika, V. K. Mulikb, P. M. Prajapatic, B.S. Shivasankard, S.V. Suryanarayanae, K.C. Jagadeesanf, S.V. Thakaref, S.C. Sharmae, A. Goswamia, Nucl. Phys. A 913 (2013) 185.

\bibitem{Liu2006}M. Liu, N. Wang, Z.X. Li, X.Z. Wu, E.G. Zhao, Nucl. Phys. A 768 (2006) 80.

\bibitem{Wang2009}N. Wang, M. Liu, Y.X. Yang, Sci. China Ser. G 52 (2009) 1554.

\bibitem{P.W.1988}P.W. Lisowski, J.L. Ullman, S.J. Balestrini, A.D. Carlson, O.A. Wasson, N.W. Hill, Neutron induced fission cross section ratios for $^{232}$Th, $^{235}$, $^{238}$U, $^{237}$Th and $^{239}$Pu from 1 to 400 MeV, Report 1988, 88MITO, 97.

\bibitem{Paradela2006}C. Paradela, L. Audouin, n-TOF fission data of interest to GEN-IV and ADS, 2006VANCOU/B076, 2006, or 2007NICE/1-421, 2007.

\bibitem{Shcherbakov2001}O. Shcherbakov, A. Donets, A. Evdokimov, A. Fomichev, T. Fukahori, A. Hasegawa, A. Laptev, V. Maslov, G. Petrov, S. Soloviev, Yu. Tuboltsev, A. Vorobyev, J. Nucl. Sci. Tech., Suppl. 2, 1 (2002) 230.

\bibitem{Koning2001}M.C. Duijvestijn, A.J. Koning, F.-J. Hambsch, Phys. Rev. C 64 (2001) 014607.

\bibitem{Seeger1976}P. Seeger, W. Howard, At. Data Nucl. Data Tables 17 (1976) 428.

\bibitem{Dobrowolski2007}A. Dobrowolski, K. Pomorski, and J. Bartel, Phys. Rev. C 75 (2007) 024613, Phys. Scr., T 125 (2006) 188.

\end{thebibliography}



\end{document}